\documentclass[twocolumn]{aastex631}

\shorttitle{
MI-driven MR in Collisionless Relativistic Jets}
\shortauthors{Kawashima et al.}
\graphicspath{{./}{figures/}}

\begin{document}

\title{
Mushroom-instability-driven Magnetic Reconnections in Collisionless Relativistic Jets
}

\correspondingauthor{Tomohisa Kawashima}
\email{kawshm@icrr.u-tokyo.ac.jp}

\author[0000-0001-8527-0496]{Tomohisa Kawashima}
\affiliation{Center for Computational Astrophysics, National Astronomical Observatory of Japan, National Institutes of Natural Sciences, 2-21-1 Osawa, Mitaka, Tokyo 181-8588, Japan}
\affiliation{Institute for Cosmic Ray Research, The University of Tokyo, 5-1-5 Kashiwanoha, Kashiwa, Chiba 277-8582, Japan}

\author[0000-0001-5804-2930]{Seiji Ishiguro}
\affiliation{Fundamental Physics Simulation Research Division, National Institute for Fusion Science, National Institutes of Natural Sciences, 322-6 Oroshi-cho, Toki, 509-5292, Japan}
\affiliation{Department of Fusion Science, 
The Graduate University for Advanced Studies, SOKENDAI,
322-6 Oroshi-cho, Toki 509-5292, Japan}


\author[0000-0003-2513-0453]{Toseo Moritaka}
\affiliation{Fundamental Physics Simulation Research Division, National Institute for Fusion Science, National Institutes of Natural Sciences, 322-6 Oroshi-cho, Toki, 509-5292, Japan}
\affiliation{Department of Fusion Science, 
The Graduate University for Advanced Studies, SOKENDAI, 
 322-6 Oroshi-cho, Toki 509-5292, Japan}

\author[0000-0002-4040-2747]{Ritoku Horiuchi}
\affiliation{Fundamental Physics Simulation Research Division, National Institute for Fusion Science, National Institutes of Natural Sciences, 322-6 Oroshi-cho, Toki, 509-5292, Japan}

\author[0000-0003-2726-0892]{Kohji Tomisaka}
\affiliation{Division of Science, National Astronomical Observatory of Japan, National Institutes of Natural Sciences, 2-21-1 Osawa, Mitaka, Tokyo 181-8588, Japan}



\begin{abstract}
We study the kinetic plasma dynamics in collisionless relativistic jets with velocity shear, by carrying out  particle-in-cell simulations in the transverse plane of a jet.
It is discovered that intermittent magnetic reconnections (MRs) are driven by Mushroom instability (MI), which is an important kinetic-scale plasma instability 
in the plasma shear-flows with relativistic bulk speed.
We refer to this sequence of kinetic plasma phenomena as "MI-driven MR".
The MI-driven MRs intermittently occur with moving the location of the reconnection points from the vicinity of the initial velocity-shear surface towards the center of the jet.
As a consequence, the number density of high energy electrons, which are accelerated by MI-driven MRs, increases with time in the region inside the initial velocity-shear surface with accompanying the generation and subsequent amplification of magnetic fields by MI.
The maximum Lorentz factor of electrons increases with initial bulk Lorentz factor of the jet.
A possible relation of MI-driven MR to the bright synchrotron emission in jet-spine of active galactic nucleus jets is also discussed.
\end{abstract}



\section{Introduction} \label{sec:intro}

The relativistic jets launched from the supermassive black holes \citep{2019ARA&A..57..467B} show or imply the enigmatic sub-structure inside them.
The elliptical galaxy M87 is one of the best laboratories to study the relativistic jet because it possesses one of the closest jets, so that its jet structure is best resolved \citep{2011Natur.477..185H,2018ApJ...855..128W}.
Recently, 
very long baseline interferometer (VLBI) observation in radio band
 revealed the triple-ridge structure in 
M87 jet at a distance of $r {\gtrsim} 10^3 r_{\rm g}$ from the central supermassive black hole  \citep{2016ApJ...833...56A,2016A&A...595A..54M,2017Galax...5....2H}, where the distance is projected on the sky plane.
Here, $r_{\rm g} = GM/c^2$ is the gravitational radius of the central black hole with its mass $M$,
$G$ is the gravitational constant, 
and $c$ is the speed of light.
The triple-ridge structure can be interpreted as the observational feature of the spine-sheath structure of the jet, which is  composed of the fast flowing spine and the slow flowing sheath
\citep{2002MNRAS.336..328L}.
The spine-sheath structure is also proposed to explain the short-term variation of TeV $\gamma$-ray emission in 
blazars (e.g., Mrk 501 and PKS 2155-304) and may be  ubiquitous in the relativistic jet \citep{2005A&A...432..401G}.

In order to understand the physics and the observed features of the relativistic jets, a number of  magnetohydrodynamic (MHD)  simulations have been performed 
\citep{1999ApJ...522..727K,2006MNRAS.368.1561M, 2007ApJ...662..835M,  2011MNRAS.418L..79T, 
2015MNRAS.452.1089P, 2018NatAs...2..167G, 2018ApJ...868..146N, 2019A&A...632A...2D, 2019MNRAS.490.4271M}. 
The general relativistic radiation MHD (GRRMHD) simulations qualitatively reproduced the sheath brightened structure \citep{2019MNRAS.486.2873C}. In addition, a possible formation of the plasmoids in the sheath is proposed by general relativistic MHD simulations \citep{2018ApJ...868..146N} and general relativistic resistive MHD simulations \citep{2020ApJ...900..100R}.
However, GR(R)MHD simulations could not explain the formation of the jet spine up to today\footnote{Alternatively, it has long been known that the natural place for under expanded hypersonic jets (in rocket jets, in particular) to dissipate is at the center line. Particles are energized downstream of Mach disks (shocks) that form along the center-line due to compressive waves emanating from the boundary \citep{1983ApJ...266...73S,2014PhFl...26i6101E}.}.
Meanwhile, 
a limited number of  particle-in-cell (PIC) simulations of jets have been carried out as a strong tool to explore the 
kinetic plasma 
processes of jets \citep{2003ApJ...595..555N, 2016ApJ...820...94N, 2018PhRvL.121x5101A,2020ApJ...896L..31D}. 
In the 2010s, an important electron-scale shear-instability called Mushroom instability \citep[MI;][]{2012ApJ...746L..14A} was found in the plasma with relativistic bulk shear flow.
Importantly, the MI dominates the electron-scale Kelvin-Helmholtz instability 
when relative bulk speed is greater than ${\sim} 0.3c$ \citep{2015PhRvE..92b1101A}. 
Based on kinetic plasma simulations \citep{2012ApJ...746L..14A, 2013PhRvL.111a5005G, 2013ApJ...766L..19L, 2014NJPh...16c5007A, 2015PhRvE..92b1101A}, the mechanism of MI is the following:
At first, the interchanging thermal motion of electrons across the shear surface, which is faster than proton motions, results in the electric currents along the jet bulk-motion near the shear surface.
Then, the magnetic field is  generated mainly parallel to the shear surface. 
The generated magnetic field further induces the interchanging electron motion across the velocity-shear surface, and the perturbation grows into a number of electron channel-flows.

\begin{figure*} [!ht]
\centering
\includegraphics[width=0.99\textwidth]{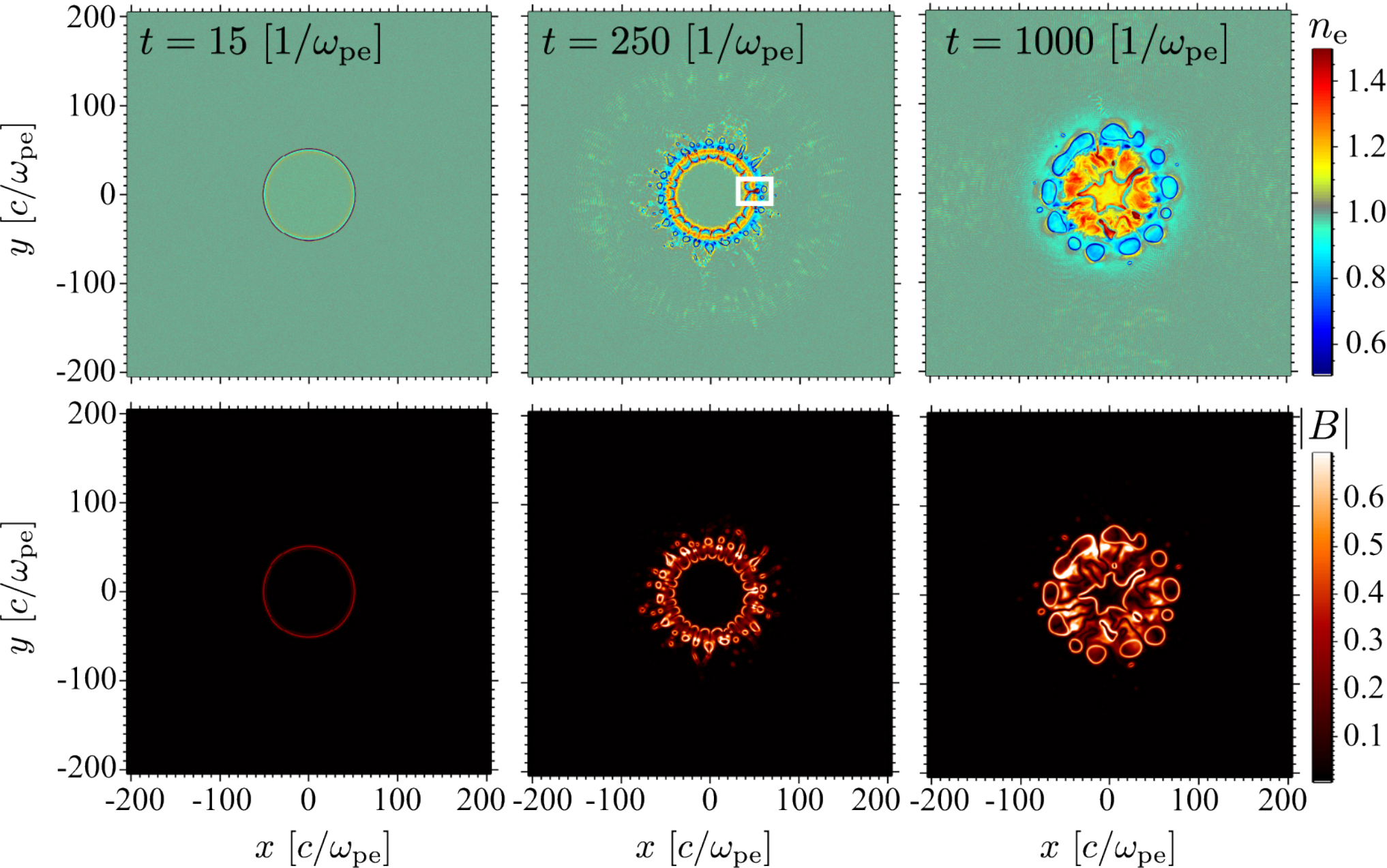} 
\caption{
Time sequence of PIC simulations in the entire simulation box. The top and bottom panels display the electron number density and the 
magnetic field strength, respectively. 
The white rectangular overlaid in the top middle panel represents the region of the magnified view shown in Figure \ref{fig:MI-driven-MR}.
The physical quantities in the figures are displayed  with an arbitrary unit.
}
\label{fig:overview_fiducial}
\end{figure*}

In this paper, we study the kinetic plasma dynamics triggered by the MI, namely Mushroom-instability-driven magnetic reconnection (MI-driven MR), by carrying out two-dimensional particle-in-cell (PIC) simulations on the transverse plane of the relativistic jets with velocity-shear surface.
Here, MR is the topological rearrangement of magnetic fields with dissipation of magnetic energy and resultant heating and acceleration of plasma particles \citep{1979cmft.book.....P, 1999PhPl....6.4565H, 2000mrp..book.....B, 2001ApJ...562L..63Z, 2006Natur.443..553D, 2010RvMP...82..603Y, 2014ApJ...783L..21S}.
We demonstrate that the intermittent MI-driven MR results in the electron concentration to the center (i.e., spine).
We propose that this new kinetic plasma dynamics will lead to the  formation of the jet spine in the relativistic jets.

\section{Model Setup} \label{sec:floats}

We perform fully kinetic PIC simulations of the relativistic jets of electron-proton
plasma. 
The simulations are carried out by using a two-dimensional relativistic PIC simulation code \texttt{PASTEL} \citep{2016PhPl...23c2110M} in the Cartesian coordinate system with
$x$--$y$ plane, which can explicitly solve the Maxwell equations and the Newton-Lorentz equation in a self-consistent way \citep[e.g.,][]{1991ppcs.book.....B, 1988csup.book.....H,2001CoPhC.135..144E}.
We set the $x$--$y$ plane in the transverse direction of the jet, i.e., the propagation direction of the jet is perpendicular to the computational domain.
Although the computational domain is in two-dimension, the physical quantities have three dimension in space, i.e., $z$-component of physical quantities of the plasma and electromagnetic fields are also time integrated.

The size of the simulation domain is $(409.6 \times 409.6)$, where the unit of the length is the electron skin depth $c/\omega_{\rm pe}$, where $\omega_{\rm pe}$ is the plasma frequency of electrons.
This simulation domain is resolved by $(8192 {\times} 8192)$ grid cells, and the cell size is one-half of the Debye length of the electrons.
The periodic boundary condition is imposed at each boundary of the computational domain.

The initial condition of the simulation is as follows.
We define the cylindrical radius $r = (x^2 + y^2)^{1/2}$ and that of the initial velocity-shear surface $r_{\rm shear} = 50 c/\omega_{\rm pe}$. 
We set the high-bulk velocity plasma with the speed $v_{\rm bulk}$ whose corresponding Lorentz factor is $\Gamma_{\rm bulk}$ in the region $r \le r_{\rm shear}$.  
We note that, in this work,
we assume the bulk speed to be zero in the region $r > r_{\rm shear}$ to focus on kinetic plasma phenomena in a simplified model setup, rather than setting a structured jet with velocity shear inside a static ambient plasma, which may be realized in a more realistic situation.
We examine a moderately relativistic bulk velocity model  with $v_{\rm bulk} = 0.9 c$ ($\Gamma_{\rm bulk} \simeq 2.3$) and a highly relativistic bulk velocity model with $v_{\rm bulk} \simeq 0.99 c$ (i.e., $\Gamma_{\rm bulk} = 10$).
The temperatures of the electrons and protons are constant in space, and 
their thermal velocities equal to $0.1 c$ and $0.025 c$, respectively. 
The particles are also uniformly distributed and neither magnetic field nor electric current is assumed.
We add no artificial initial perturbation to the plasma.
The adequacy of our simple assumption on no initial magnetic field, which is expected to be consistent for the jet plasma far from the central black hole, will be discussed in \S \ref{sec:summary}. 

\begin{figure*}[!ht]
\begin{center}
\includegraphics[width=0.9\textwidth]{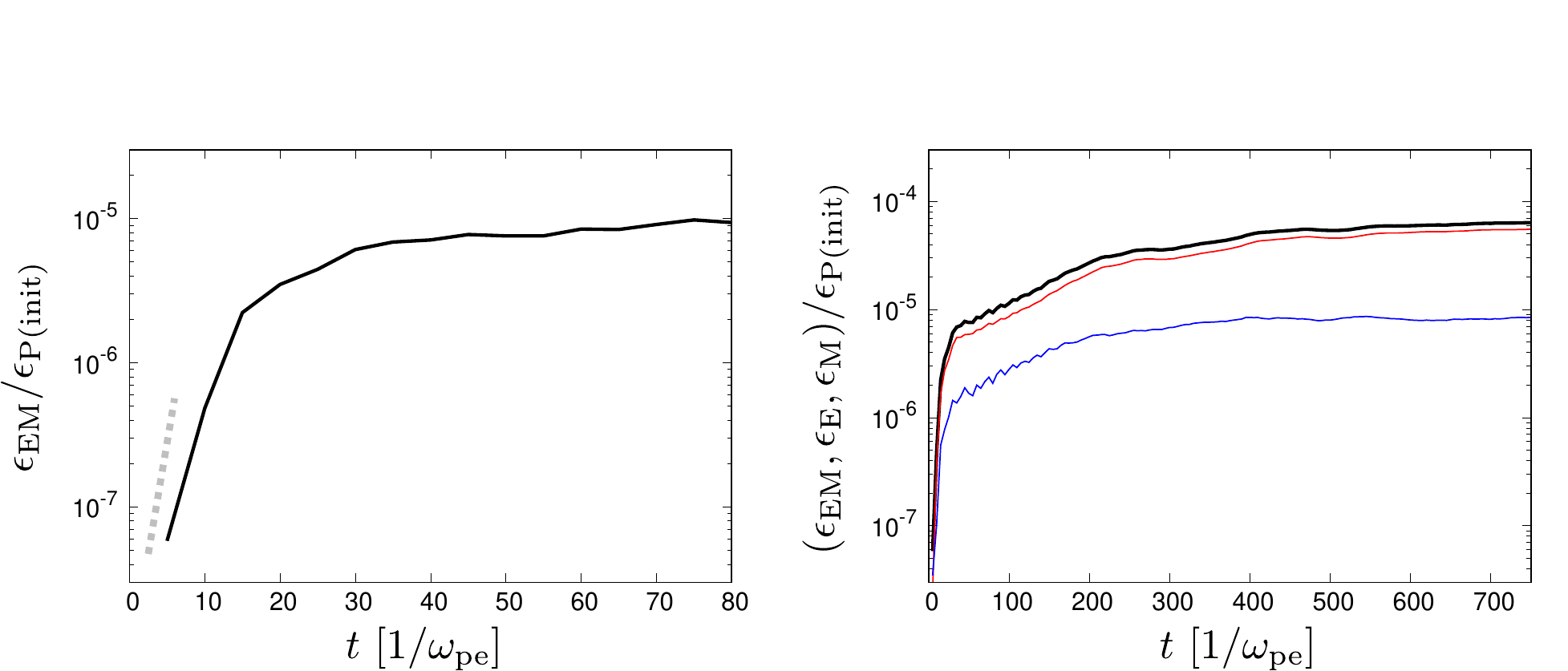}
\caption{
Time evolution of energy of electromagnetic field $\epsilon_{\rm EM}$, electric field $\epsilon_{\rm E}$, magnetic field $\epsilon_{\rm M}$ normalized by initial kinetic energy of particles (i.e., protons and electrons) $\epsilon_{\rm P(init)}$, which are measured in the laboratory frame. The left panel shows the early stage of simulation. The black curve represents $\epsilon_{\rm EM}$.
As a reference, the slope of maximal growth rate of MI ($\propto {\bar v}/c \sqrt{{\bar \gamma}}$) for cold plasmas is shown by the gray dashed line. The MI grows exponentially up to $t~\sim ~10/\omega_{\rm pe}$.
The right panel shows the long term evolution of $\epsilon_{\rm EM}$ (black), $\epsilon_{\rm E}$ (blue), and $\epsilon_{\rm M}$ (red).}
\label{fig:time_energy}
\end{center}
\end{figure*}

Since the injection mechanism of the plasmas into the jet is an open question, we simply assumed the initially uniform density plasma composed of protons and electrons as described above. 
The protons and electrons might be injected into the jet base as a result of the MRs triggered at the interface of jet and turbulent accretion flow
\citep{2011MNRAS.418L..79T} or the decay of the neutrons generated via the accelerated proton in the underlying accretion flows \citep{2012ApJ...754..148T}.

\begin{figure*}[!ht]
\begin{center}
\includegraphics[width=0.95\textwidth]{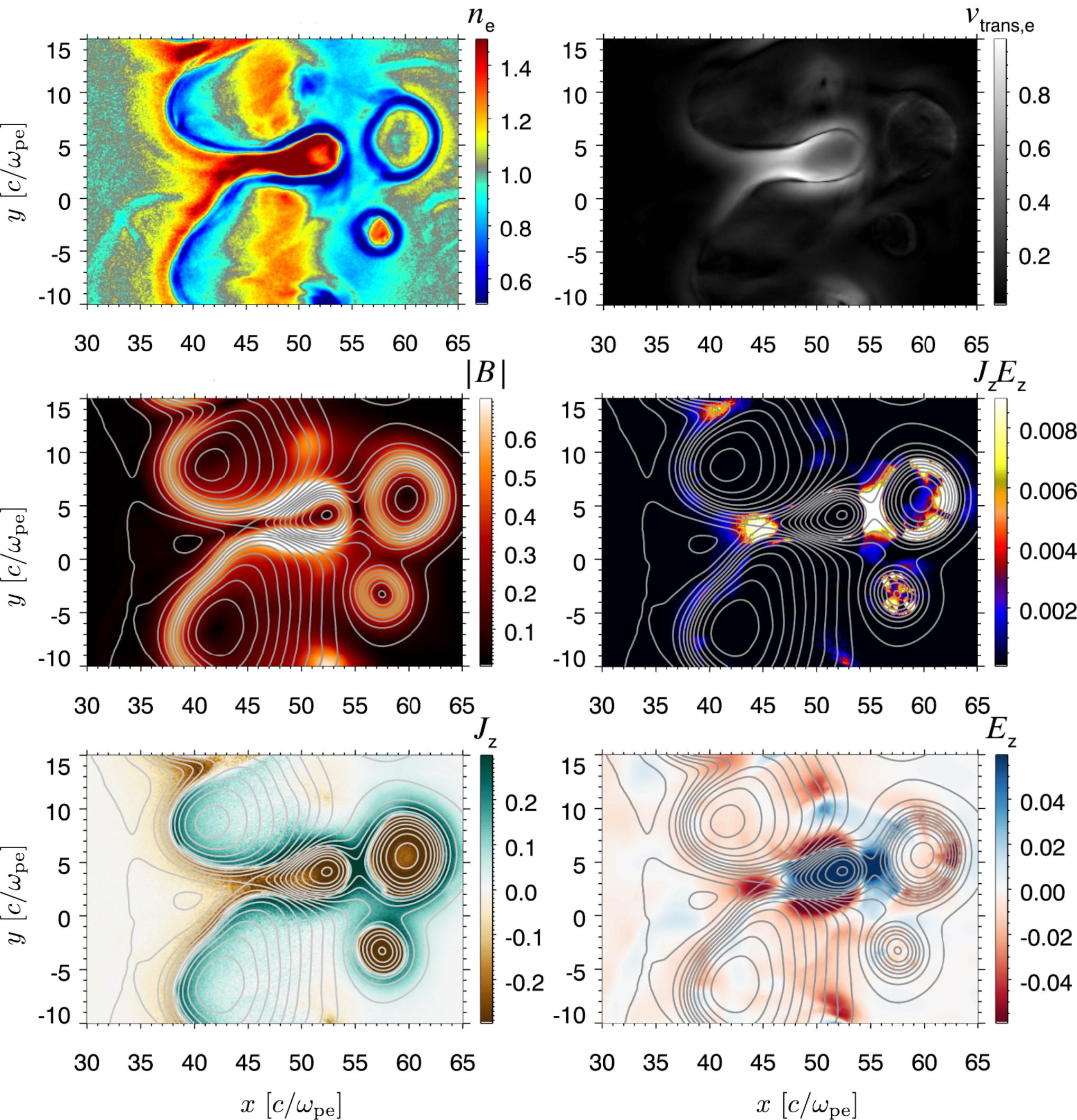} %
\caption{
Magnified view of the electron density (top left),  
the absolute value of the electron four-velocity 
in the transverse direction of the jet
(top right), the magnetic field strength (middle left), 
the product of $z$ component of electric current density and electric field (middle right), $z$ component of electric current density (bottom left), and $z$ component of electric field (bottom right) in the region enclosed in Fig. 1 at $t = 2.5 \times 10^2 ~ \omega_{\rm pe}^{-1}$.
The electric current and the electric field are evaluated in the co-moving frame of the electrons at each point.
The gray lines in the middle and bottom panels display the magnetic field lines in the plane.
}
\label{fig:MI-driven-MR}
\end{center}
\end{figure*}

\begin{figure} [!h]
\includegraphics[width=0.99\columnwidth]{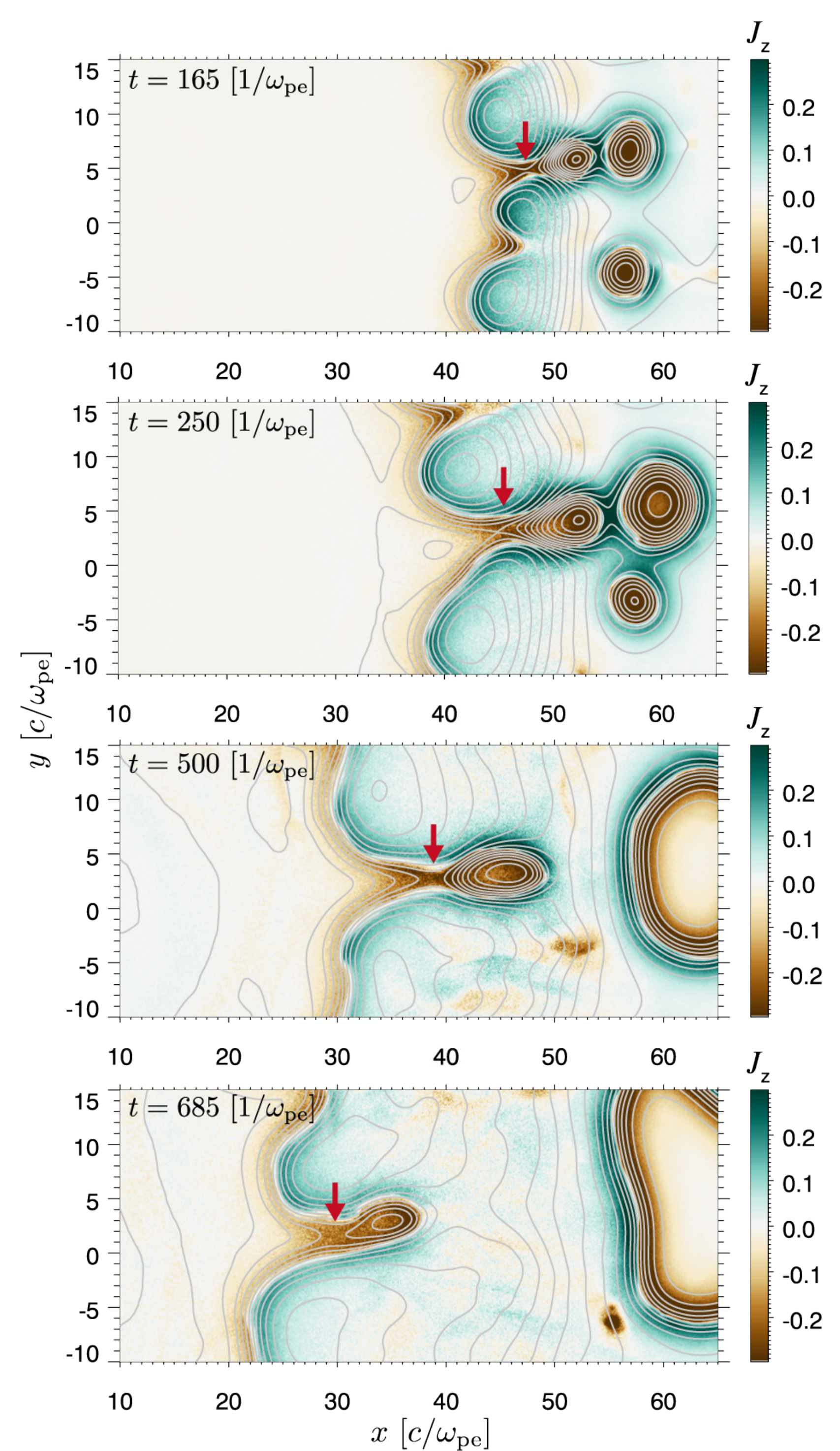} 
\caption{Time evolution of the intermittent MI-driven MRs near the velocity-shear surface. The electric current sheet and the reconnection points (red arrows) shift towards the jet center with time.
The electric current is evaluated in the co-moving frame of electrons at each point. 
}
\label{fig:time_MI-driven-MR}
\end{figure}

\section{Results} 
\subsection{Moderately Relativistic Bulk Speed Model ${\rm (} v_{\rm bulk} = 0.9c , {\rm i.e.}, \Gamma_{\rm bulk} \simeq 2.3 ${\rm )}}

\begin{figure}[!h]
\begin{center}
\includegraphics[width=0.45\textwidth]{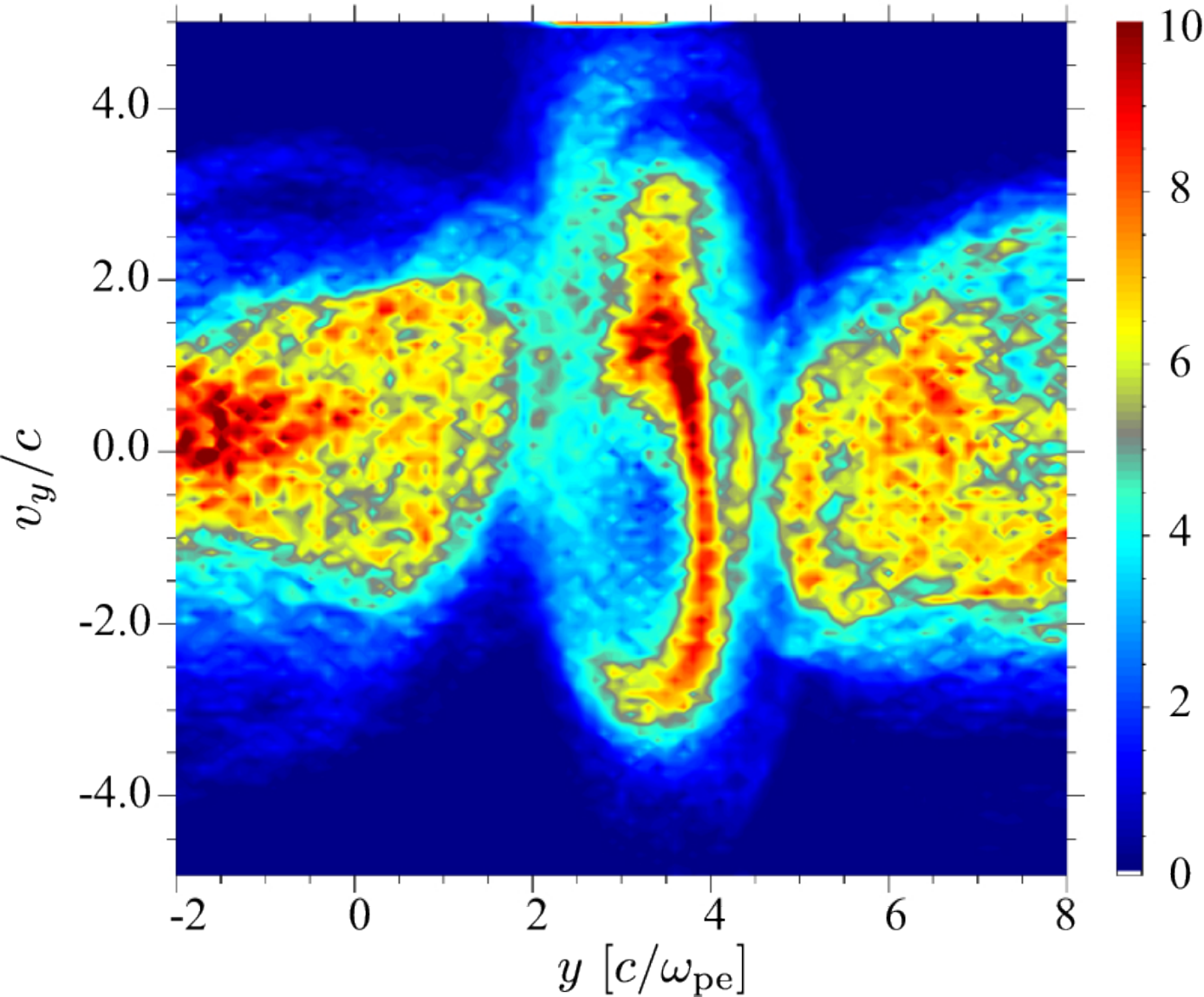} 
\caption{
The electron distribution in $y$--$v_y$ phase space at $t = 2.5 \times 10^2 \omega_{\rm pe}^{-1}$, in which electrons are sampled along the line $x = 46 c/\omega_{\rm pe}$. The hole structure at $(y, v_{y}/c) {\simeq} (3 c/\omega_{\rm pe}, 0)$ evidently presents the appearance of the MR at this point.
In this figure, the velocity in $y$-direction is evaluated in the co-moving frame of the bulk motion in $z$-direction.}
\label{fig:MR-hole}
\end{center}
\end{figure}

\begin{figure*}[!ht]
\begin{center}
\includegraphics[scale=0.8]{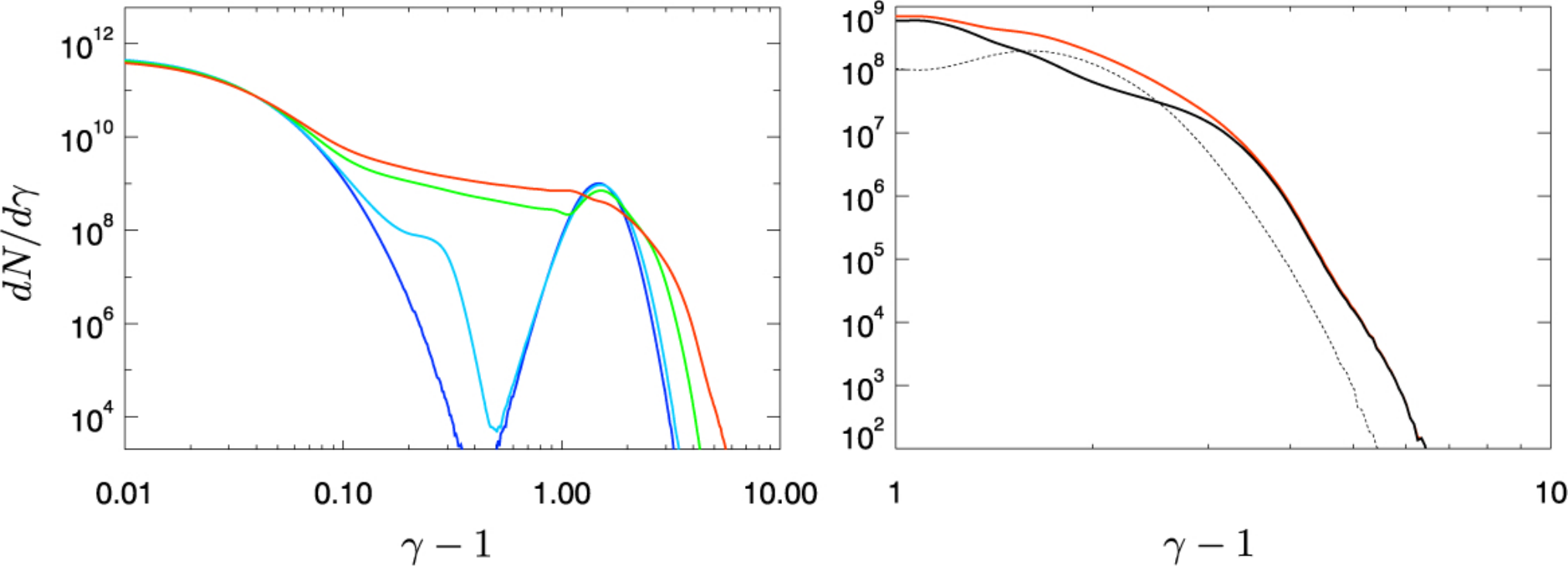} 
\caption{
Left: Time evolution of the number spectra of the electrons at $t = 0$ (blue), 20 (cyan), 250 (green), and $750 ~ \omega_{\rm pe}^{-1}$ (red).  Right: energy spectra of electrons at $t = 750 ~ \omega_{\rm pe}^{-1}$ in the range $1 \le \gamma - 1 \le 10$. The spectrum of electrons in the whole simulation domain (red curve) is decomposed into those of electrons which currently exist inside/outside the initial shear surface (black solid-thick/dotted-thin curve).}
\label{fig:spectrum_fiducial}
\end{center}
\end{figure*}

Figure \ref{fig:overview_fiducial} demonstrates the overview of the time sequence of our PIC simulation in the entire simulation box. 
Although the magnetic field is initially zero, the MI generates and subsequently amplifies the magnetic field (bottom panels), as shown in the next paragraph in detail.
The electrons concentrate in the center of the jet region, as the relativistic jet propagates (top panels).
This is due to the combination of magnetic pinching force induced by MI and the high energy electron ejections by the subsequent, intermittent MRs. 
We refer to this new type MR as MI-driven MR.
The electrons are accelerated via the MRs.
Since the high energy electrons concentrate in the jet-center region together with amplified magnetic field, the jet-center is expected to be bright in the synchrotron emission as discussed in \S \ref{sec:summary}.


We overview the time evolution of MI in the linear-stage and subsequent MI-driven MRs in the non-linear stage. 
At the early stage of our simulation (up to $\sim 15/\omega_{\rm pe}$), the MI exponentially grows near the velocity-shear surface as follows:
(1) the thermal electron motion across the velocity-shear surface leads to the generation of
electric current in $z$ direction in the vicinity of the velocity-shear surface.
(2) The magnetic fields surrounding the jet center (i.e., high-bulk-Lorentz-factor region) are subsequently generated near the shear surface (left-bottom panel of Figure \ref{fig:overview_fiducial}). 
(3) Then, the electrons experience the Lorentz force, which 
enhances the electron motion to cross the shear surface. 
This is the linear stage of MI and the timescale of the instability $\sim~1/\omega_{\rm pe}$ is consistent with the growth timescale of MI assuming cold plasma obtained by the linear analysis 
$t_{\rm MI} ~\sim~1/\omega_{\rm pe}$
\citep[][see also Appendix]{2015PhRvE..92b1101A}.
The left panel of Figure \ref{fig:time_energy} shows that  the MI exponentially grows in this timescale and the growth rate of the electromagnetic energy is consistent with the maximum linear growth rate obtained by the linear analysis of the MI.

In the non-linear stage of the MI,
the channel flow is evidently formed into the direction parallel to the shear surface, as a consequence of the earlier linear growth \citep{2015PhRvE..92b1101A}. 
The MR is
recurrently induced in this channel-flow region (that is,  MI-driven-MR).
The channel flow of electrons in the radial direction, which is induced by the Lorentz force due to the  magnetic field generated in the azimuthal direction, 
accompanies the strong electric current in the $z$ direction and the anti-parallel configuration of the magnetic field elongated in the radial direction (Figure \ref{fig:MI-driven-MR} as discussed later). 
Then, the MR is triggered near the shear surface and the heated and/or accelerated electrons are ejected towards the regions inside and outside the shear surface as shown in the next paragraph.
At the early non-linear stage ($t ~\lesssim~ 200 \omega_\mathrm{pe}^{-1}$), the electromagnetic energy intermittently varies. 
This is due to the episodic MI-driven MR; the MRs convert the electromagnetic energy amplified by the MI to the electron energy, while the MI recovers the magnetic energy lost by the MRs and forms new electric current sheets followed by subsequent MRs.
We note that the MI-driven MR continues to occur, although the time variation of the electromagnetic energy in the whole simulation domain is not remarkable after the early non-linear stage.
In the later non-linear stage of MI ($t ~\gtrsim~ 400 \omega_\mathrm{pe}^{-1}$), the electromagnetic energy is amplified up to $10^{-4}$ of the initial particle energy in the whole simulation domain.
Throughout the simulation, the magnetic energy dominates the electric energy since the simulated plasma is mildly relativistic.
These are shown in the right panel of Figure \ref{fig:time_energy}.

Next, we present the detail of MI-driven MR near the velocity-shear surface.
Figure \ref{fig:MI-driven-MR} displays magnified view of the electron density, transverse electron velocity, magnetic field strength, energy conversion rate from the electromagnetic field  to electrons, electric current density, and electric field 
at  $t = 2.5 \times 10^2\, \omega_{\rm pe}^{-1}$.
As a result of the nonlinear evolution of the MI, an elongated current sheet surrounded by  anti-parallel magnetic fields grows in the radial direction (see $(x, y) ~ \sim ~ (45, 3)c\omega_\mathrm{pe}^{-1}$ in the bottom-left panel). The direction of the electric current density is anti-parallel to the bulk-jet motion, since the electric current is generated via the MI. Importantly, X-shaped singular points of magnetic field lines ("reconnection points") appear at the center of the current sheet, in which the reconnection component ($z$-component) of electric field is generated along the electric current density  through a microscopic kinetic process \citep{2005PhRvL..95d5003I}, and thus the field energy is strongly converted to electrons.  
In other words, the subsequent heating and acceleration of the electrons occur via the MI-driven MR.
We note that the acceleration of electrons to the highest energy in our simulation is due to the electric field of MRs.

The high-energy electrons concentrate to the region inside the initial velocity-shear surface as the intermittent MI-driven MRs evolve. 
In Figure \ref{fig:time_MI-driven-MR}, we show time evolution of the episodic MI-driven MRs. One can find that the reconnection point moves towards the jet center as time increases, which results in the contraction of the current sheet surrounding the jet center. 
This is because the electrons which generate the electric current move towards the jet center due to the MRs and the Lorentz force (the magnetic pinch force)  as a consequence of MI.
As the current sheet  contracts, the reconnection point moves towards the jet center. Then, the accelerated and/or heated electrons concentrate in the jet center, i.e., the jet spine is formed.

Here, we confirm that the MRs are physically triggered in our simulation.
Figure \ref{fig:MR-hole}
displays a phase-space ($y$--$v_{y}$) plot of electrons at $t=250 \omega_{\rm pe}^{-1}$ in a region where an MR expected to be taking place at $x = 46 c/\omega_{\rm pe}$, i.e., the region near the red arrow in the second top panel in Figure \ref{fig:time_MI-driven-MR}.
One can clearly find the appearance of the hole structure at around $y ~{\simeq}~ 3 c/\omega_{\rm pe}$ and $v_{y}~ \simeq ~0$.
This hole structure in the phase-space is a consequence of the 
stochastic motion of electrons in the vicinity of the magnetic neutral sheet, i.e., meandering motion \citep{1965JGR....70.4219S}. 
The hole structure in the phase space is
a strong evidence of the MRs \citep{HO2008}.

The MI-driven-MRs accelerate electrons.
Figure \ref{fig:spectrum_fiducial} 
displays the time evolution of electron energy-spectra in the whole computational domain.
Initially, the electrons outside the velocity-shear surface shows the Maxwell distribution, while the shifted-Maxwellian appears inside the shear surface because the electrons have the bulk velocity of $v=0.9c$.
One can find that the more electrons are accelerated with time via MI-driven MR and the power-law spectra with narrow energy range appear as explained below.

Just after the MI grows, the electrons outside the shear surface are mainly accelerated and the
Maxwell distribution starts to be distorted around  $\gamma - 1 ~ \sim 0.1$ at $t~{\gtrsim} ~ 20~ \omega_{\rm pe}^{-1}$.
The MRs intermittently take place with their reconnection points, which move towards the center of the cylindrical jet with time as shown in Figure \ref{fig:time_MI-driven-MR}.
At $t = 750 ~\omega_{\rm pe}^{-1}$, 
the power-law spectra with narrow energy range is formed at $\gamma - 1 \simeq 1$. 
As shown in the right panel of 
Figure \ref{fig:spectrum_fiducial}, 
the highly accelerated particles are distributed inside the initial shear surface, indicating that the jet spine is composed of high energy electrons.
The high energy electrons are mainly accelerated to the jet direction ($z$-direction) by virtue of the electric fields of MI-driven MRs.
The maximum Lorentz factor of electrons reaches $\sim 8$ at the end of our simulations in this model.

\subsection{Highly Relativistic Bulk Speed Model ${\rm (} v_{\rm bulk} \simeq 0.99c , {\rm i.e.}, \Gamma_{\rm bulk} = 10 {\rm )}$}

\begin{figure*} [!ht]
\centering
\includegraphics[width=0.9\textwidth]{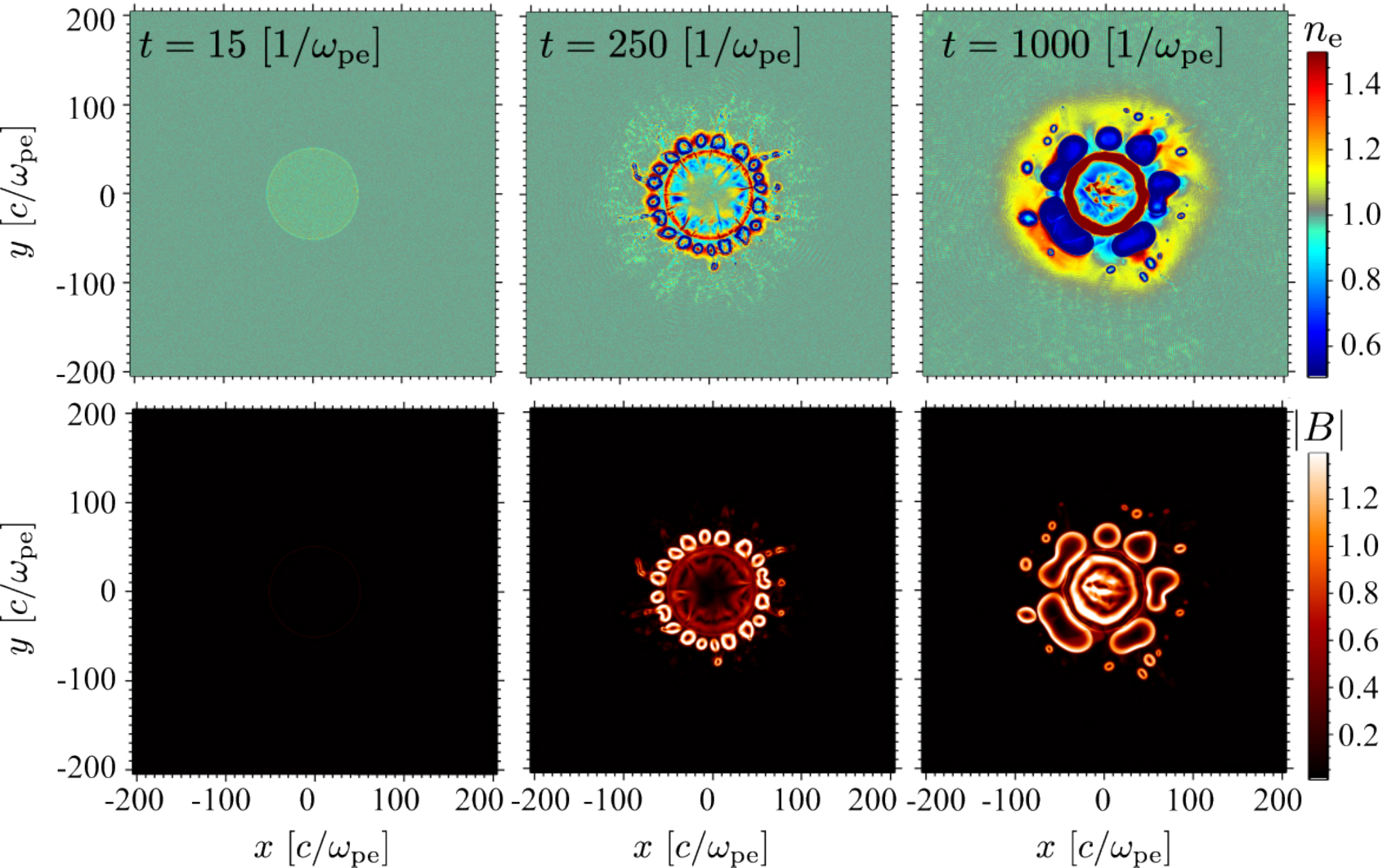} 
\caption{
Same as Fig.\ref{fig:overview_fiducial} but for highly relativistic bulk speed model $( v_{\rm bulk} \simeq 0.99c , {\rm i.e.}, \Gamma_{\rm bulk} = 10)$.}
\label{fig:overview_high}
\end{figure*}

A model with highly relativistic bulk speed is  examined here.
Figure \ref{fig:overview_high} displays the time evolution of the MI-driven MRs in the plasma with high velocity shear.
As in the case of moderately relativistic bulk speed model, the MI-driven MR occurred.
The highly relativistic shear flow brings much stronger magnetic field via the MI compared with the moderately relativistic former model.
The electron number density in a part of the high-bulk-Lorentz-factor region ($20 c/\omega_{\rm pe} \lesssim r \lesssim 40 c/\omega_{\rm pe}$) is small, while it is high in the rest of the high-bulk-Lorentz-factor region (i.e., messy region in the center $r \lesssim 20 c/\omega_{\rm pe}$ and the high density ring-like region $ 40 c/\omega_{\rm pe}  \lesssim  r \lesssim  50 c/\omega_{\rm pe}$)) at $t=1000 \omega_{\rm pe}^{-1}$. 
This is due to the strongly amplified magnetic field near the strong velocity-shear region, since the stronger magnetic field results in the smaller gyro-radius of electrons and a smaller number of the electrons move towards the jet center. 
It should be noted that the generated magnetic field will be so strong that the kink instability can occur, which will be discussed in \S 4.

\begin{figure*}
\begin{center}
\includegraphics[scale=0.8]{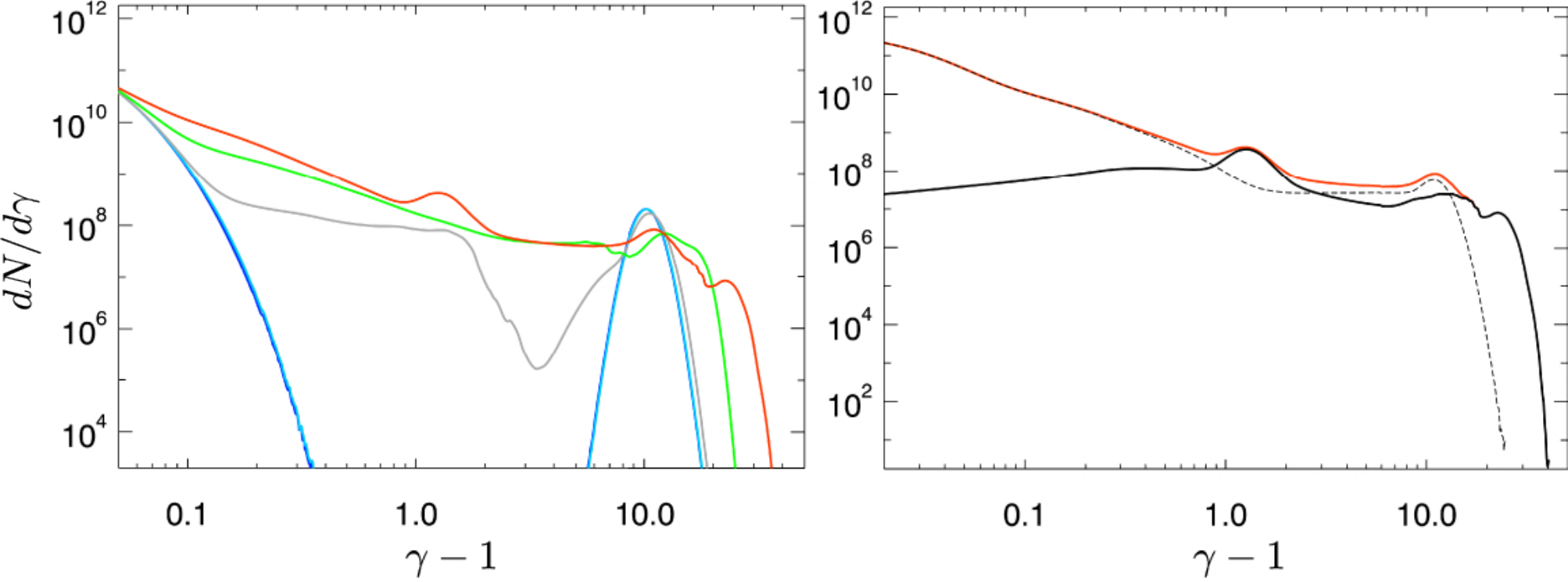} 
\caption{
Same as Fig.\ref{fig:spectrum_fiducial} but with highly relativistic bulk speed.
The time sequence of the spectra in the left panel is  $t = 0$ (blue), 20 (cyan), 70 (gray), 250 (green), and $750 ~ \omega_{\rm pe}^{-1}$ (red).
The decomposed spectrum at $t = 750 ~ \omega_{\rm pe}^{-1}$ in the right panel is shown in the range $0.02 \le \gamma - 1 \le 50$.}
\label{fig:spectrum_high}
\end{center}
\end{figure*}

Figure \ref{fig:spectrum_high} shows the time evolution of the electron spectrum. 
By virtue of the strong velocity shear, which is an energy source of the electron acceleration via MI-driven MRs, the electrons are accelerated up to $\gamma \sim 40$ (left panel), which is significantly larger than that of the mildly relativistic case, in which the electrons are accelerated up to $\gamma \sim 8$. 
The accelerated electrons exists inside the initial velocity-shear surface (right panel) as is the same as the moderately relativistic bulk velocity model.
Since we focus on the discovery of the MI-driven MR, the detailed acceleration mechanism of electrons will be left as a future work.

\section{Summary and Discussion} \label{sec:summary}

We carried out two dimensional PIC simulations to study the kinetic plasma dynamics in the transverse direction of the relativistic jet with velocity-shear surface.
Although we start the simulation with no initial magnetic field, the magnetic fields are generated by MI near the velocity-shear surface, and it is discovered that intermittent MRs induced by MI, which we refer to as "MI-driven MRs", occur in the transverse plane of the jet.
As a result, the number density of high energy electrons increases with time in the jet-center region accompanying with the generation and amplification of magnetic field by MI.
The maximum Lorentz factor of the accelerated electrons increases with the initial bulk Lorentz factor inside the velocity-shear surface.

\begin{figure}
\begin{center}
\includegraphics[width=\columnwidth]{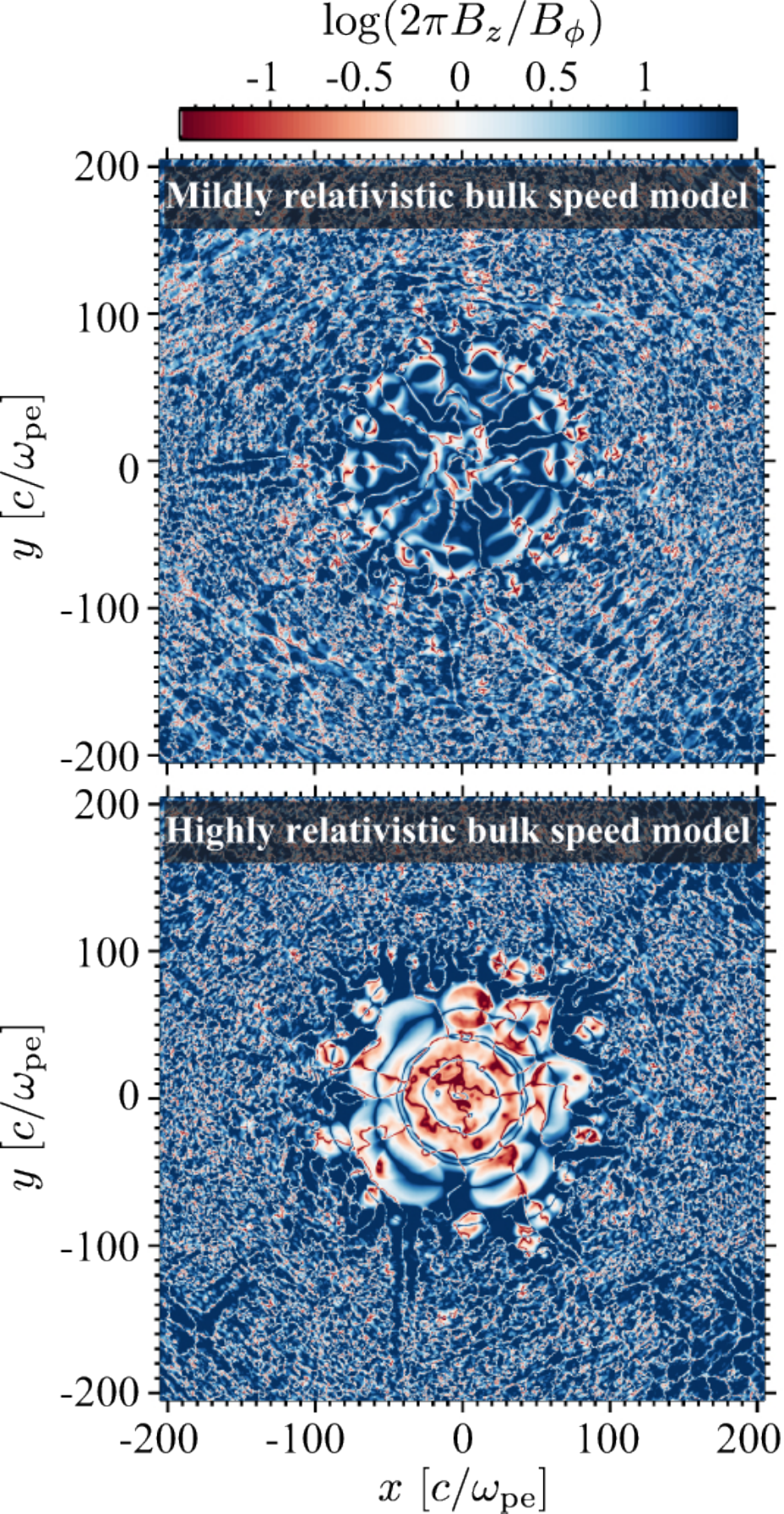} 
\caption{
Distributions of $2\pi B_z/B_\phi$ at $t= 1000 \omega_{\rm pe}^{-1}$ for mildly relativistic bulk speed model (top) and highly relativistic bulk speed model (bottom).}
\label{fig:kink_criterion}
\end{center}
\end{figure}

\begin{figure}[!h]
\centering
\includegraphics[width=\columnwidth]{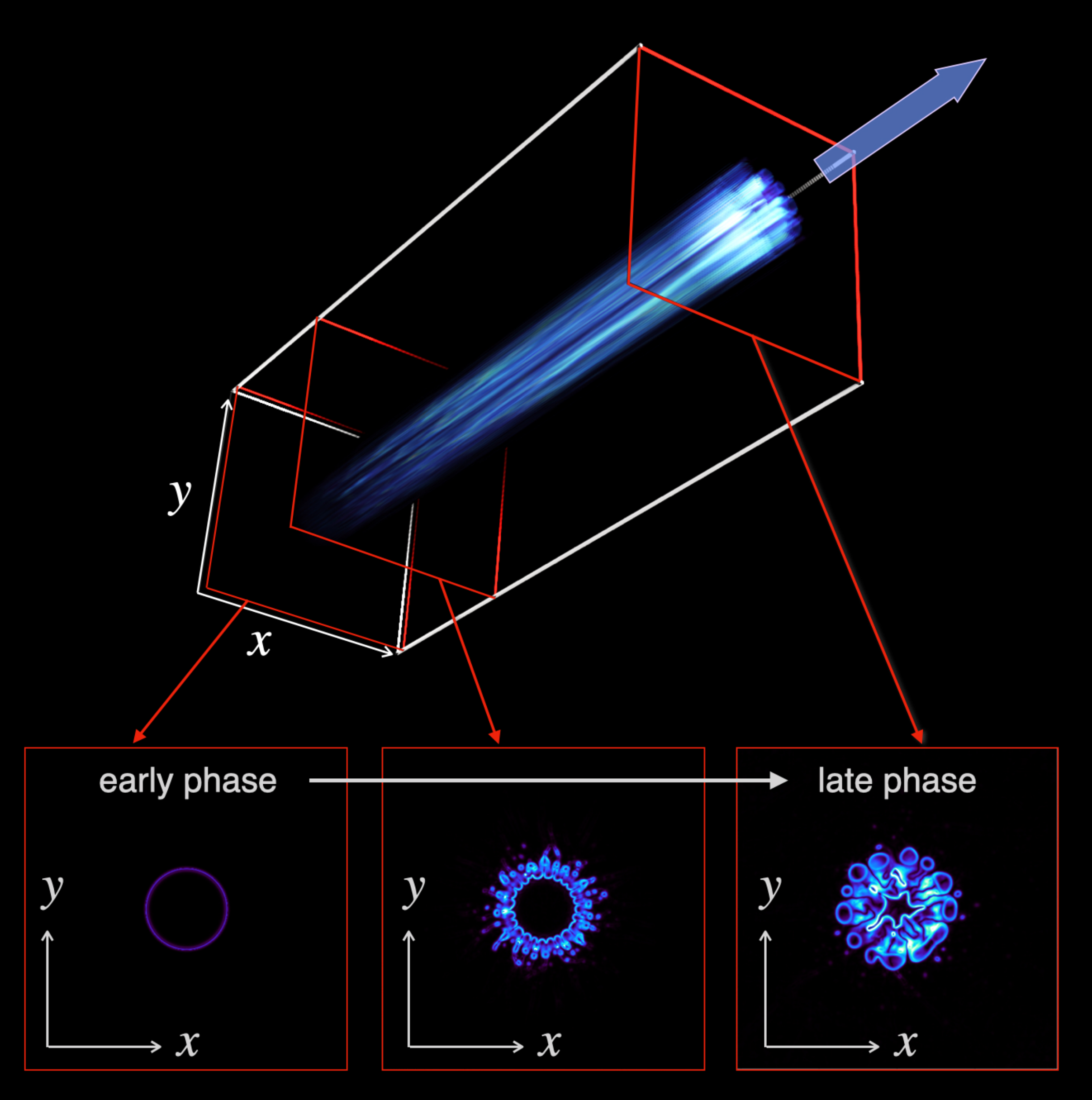} 
\caption{
Synthetic picture of our moderately relativistic jet model. 
The top panel is composed of the volume rendered image of our PIC simulation. 
 This image is synthesized by stratifying the time snapshots of our 2D PIC simulation shown in bottom panels in the direction of the jet propagation.
The color represents the electron density multiplied by the strength of the magnetic field shown in the bottom panels, which roughly agrees with the synchrotron emissivity.
In bottom panels, the epoch of the simulation is the same with that in Figure \ref{fig:overview_fiducial}, i.e., $t=$ 15, 250, and 1000 $\omega_{\rm pe}^{-1}$, from left to right.
}
\label{fig:synthetic}
\end{figure}



We discuss the consistency of our numerical setup for the dynamics of MI-driven MRs with no initial magnetic field. 
The necessary condition is that the Larmor radius of the electrons in the jets in which the MI-driven MRs are expected to take place is larger than the length of our simulation domain. 
In this case, the effects of the initial magnetic field can be reasonably neglected.
The Larmor radius of the electrons before the acceleration via the MI-driven MRs (i.e., ${\gamma}_{e} \sim 1$) is expected to be  $\gamma_e m_e c^2/e B \sim 10^{7} \gamma_e (B/1{\rm mG})^{-1}$ cm, as the magnetic field could be dissipated and its strength would be $\sim 1$ mG  \citep[e.g.,][]{2020MNRAS.492.5354M} while that near the black hole is indicated to be $\gtrsim 1$ G \citep{2021ApJ...910L..13E}.
On the other hand, the simulation box size is $\sim 100 c/\omega_{\rm pe} \gtrsim 10^6$cm.
This is because the electron skin depth in the jet can be roughly estimated to be 
$c/\omega_{{\rm p}e} = c/\sqrt{4\pi e^2 n_{\rm e}/m_{\rm e}} \gtrsim 10^4\, (n_{\rm e}/10^4{\rm cm}^{-3})^{-1/2}\,{\rm cm}$,
where  
the number density of electrons of the distant jet  will be less than that near the black hole $n_{\rm e} ~\lesssim 10^4 \,{\rm cm}^{-3}$ \citep[see, e.g.,][]{{2019ApJ...875L...5E,2021ApJ...909..168K}}.
Our simulation setup with no initial magnetic field will be reasonable, at least to study the dynamics of MI-driven MR itself,
because of $r_{{\rm L} e} \gtrsim 100\, (c/\omega_{{\rm p}e})$, 
while the PIC simulation with initial magnetic field will be performed  in forthcoming papers to confirm the results.

In the highly relativistic bulk speed model, the stronger magnetic field  surrounding the jet-center region is generated, which results in less concentration of the accelerated electrons in the jet center by suppressing the electron motion across the strong magnetic field via the Lorentz force. 
However, the strong magnetic field surrounding the high-bulk-Lorentz-factor region can be unstable against the kink instability.
Here,  we consider the  Kruskal–Shafranov stability criterion which proposes that the jet is kink unstable if $q = (2\pi r/L) (B_z / B_{\phi}) < 1$ is satisfied, where $r$,  $L$, $B_z$  and $B_\phi$ are the cylindrical radius of the jet, typical length scale of the jet in the $z$ direction, magnetic field strength in the $z$ direction, and the magnetic field strength in the azimuthal direction of the cylindrical jet, respectively.
Here, $\phi$ is the azimuthal angle in the cylindrical coordinate system defined as $\tan \phi = y/x$.
Figure \ref{fig:kink_criterion} shows the distribution of  $2\pi B_z / B_{\phi}$, which means that the plasma jet with $L > r$, which can be easily realized because of the narrow opening angle of the relativistic jets observed in active galactic nuclei, can be unstable in the region with $\log(2\pi B_z / B_{\phi}) < 0$ (i.e., red colored region).
Figure \ref{fig:kink_criterion} represents that the jetted plasma can be kink unstable for the highly relativistic bulk speed model (bottom panel), while it will be stable for the mildly relativistic bulk speed model (top panel).
In the highly relativistic bulk Lorentz factor model, strong MI growth results in the generation of intense magnetic field in the  $\phi$-direction, which suppresses the transverse motion of electrons (i.e., suppression of the generation of the magnetic field in $r$-direction). As a consequence, the electric current in $r$-direction is suppressed and the  generation of the magnetic field in the $z$-direction is reduced.
It can be proposed that the highly relativistic shear plasma flow will generate strong magnetic fields with accompanying the MI-driven MRs, and as a consequence, the large scale MRs with drastic deformation of jet structure triggered by the kink-instability and accelerate the electrons more.
After the large scale MRs, it is expected that the MI-driven MRs will restart to transport the electrons towards the jet center because a part of the strong magnetic field surrounding the jet center  will disappear just after the large scale MRs induced by the kink instability, although 3D PIC simulations are needed to confirm the expected behavior of the plasma including kink instability.

Shown in Appendix (Figure \ref{fig:MI_dispersion}), the MI will dominate the electron-scale Kelvin-Helmholtz instability in our relativistic velocity-shear flow, however, it may not be completely negligible. 
We should emphasize that the three-dimensional effects can lead to the significant destruction/suppression of the large-scale, coherent plasma structure shown in our two-dimensional simulations, since non-linear behavior of structure formation often depends on the spatial dimension
of the system.
 Three-dimensional PIC simulations to simultaneously solve the MI-driven MRs, electron-scale Kelvin-Helmholtz instability, kink instability, and others are out of the scope of this paper, so that these remain as future works.

It should be mentioned that there is 
a gap between the length-scales of PIC simulations and realistic jets in the universe.
The observed jet-spine seems to have the scale with $10^{2-3} r_{\rm g}$ $\sim$ $10^{16-17}\, (M/6.5\times10^9 M_\odot)$
cm, while the diameter of high-Lorentz-factor region in our simulation is $100 c/\omega_{{\rm p}e}$ $\sim$ $10^{6}\, (n_{\rm e}/10^4{\rm cm}^{-3})^{-1/2}$cm, i.e., the length-scale of the simulation is much smaller than that of the observed jet-spine. 
The MI-driven MR may not directly but indirectly form the jet spine via triggering an MHD scale mechanism.
Another possible scenario 
is that the MI-driven MRs continuously occur with moving their reconnection points towards inward   until the high energy electrons reaches near the jet-center (see Figure \ref{fig:time_MI-driven-MR}). 
The intermittent MRs would enable the electrons to reach the realistic jet-center because of the acceleration of the electrons and the prompt change of the magnetic field topology during MRs, while the initial and amplified magnetic field may arrest the approach of the electrons towards the jet-center.
Alternatively, 
successive shear-driven acceleration mechanism may work on the electrons after the acceleration via MI-driven MRs, and helps the electrons to penetrate the jet-center region, which is  motivated by the recent work on the shear-driven acceleration of electrons after the MRs induced by Kelvin-Helmholtz instability \citep{2021ApJ...907L..44S}. 
More realistic studies with long term PIC simulations with larger simulation box including a finite initial magnetic field and a finite width of shear-layer, which will give us more insight whether the above scenario solving the scale gap problem is correct, remain as a future work.


Here, in order to  demonstrate the possible brightening in the jet-center region as a consequence of the MI-driven MRs, Figure \ref{fig:synthetic} displays the electron  density multiplied by the  strength of the magnetic  field,  which  qualitatively  agrees  with  the  synchrotron emissivity, in the moderately relativistic bulk speed model.
One may find that the jet-center region  becomes brighter with time.
In terms of the timescale, if the straightforward extrapolation of the length scale discussed above is correct and is also applicable to non-linear structure-formation in three-dimensional plasma, the concentration of electrons in the jet spine triggered by MI is not inconsistent with the 
observation of M87. 
The typical propagation timescale of the relativistic jet up to the distance with appearance of triple-ridge structure  for M87 ($\sim 10^3 r_{\rm S}$) is $t_{\rm dyn} ~\gtrsim~ 10^3 r_{\rm S}/c ~\sim 10^8$s.
On the other hand, 
the growth timescale of MI is much shorter than the jet propagation timescale. 
In practice, the structure-formation timescale via the MI-driven MR 
should be estimated by  $r_{\rm shear}/v_{\rm trans, e}$, where $r_{\rm shear}$ is the radius of high-bulk-Lorentz-factor region and the transverse electron velocity $v_{\rm trans, e}$ is roughly equal to the initial thermal velocity of electrons.
Since the radius of M87 jet at $\sim 10^3 r_{\rm S}$ from the BH is 
$
{\sim} 100r_{\rm S}$   \citep{2012ApJ...745L..28A,2013ApJ...775..118N,2017Galax...5....2H} 
and the thermal velocity of electrons is expected to be $\gtrsim 0.1 c$, the formation timescale of the jet spine is estimated to be $\lesssim$ (several) $\times$ $10^7$s.\footnote{The important point is the ratio of the jet-spine width to the distance from the BH. If this ratio is kept, the same scenario can be applied, even if the jet-spine is found in the region closer to the BH.}
Hence, with respect to the timescale, our scenario of the jet-spine formation via MI-driven MR is consistent with the observed M87 jet.


If the MI-driven MRs take place in the relativistic jets, the azimuthal component of magnetic fields surrounding the jet spine will be generated via the MI.
The linearly polarized radio emission will be detected by VLBI observations.
The shear-driven acceleration 
through
the MI-driven MRs may also generate the VHE gamma-ray emissions.

\begin{acknowledgments}
We thank M. Kino, E. Kokubo, K. Asada, K. Hada, and M. Toida for useful discussion. This research was funded by the NINS program of Promoting Research by Networking among Institutions (Grant Number 01421701). Numerical calculation was performed on the Plasma Simulator (FUJITSU FX100) of NIFS with the support and under the auspices of the NIFS Collaboration Research program (NIFS17KNSS092). A part of the simulations were performed on the XC30 at the Center for Computational Astrophysics in NAOJ.

\end{acknowledgments}

\appendix
\section{The  maximum linear growth rate of the Mushroom instability} \label{sec:appendix}

\begin{figure}[!h]
\centering
\includegraphics[width=0.5\columnwidth]{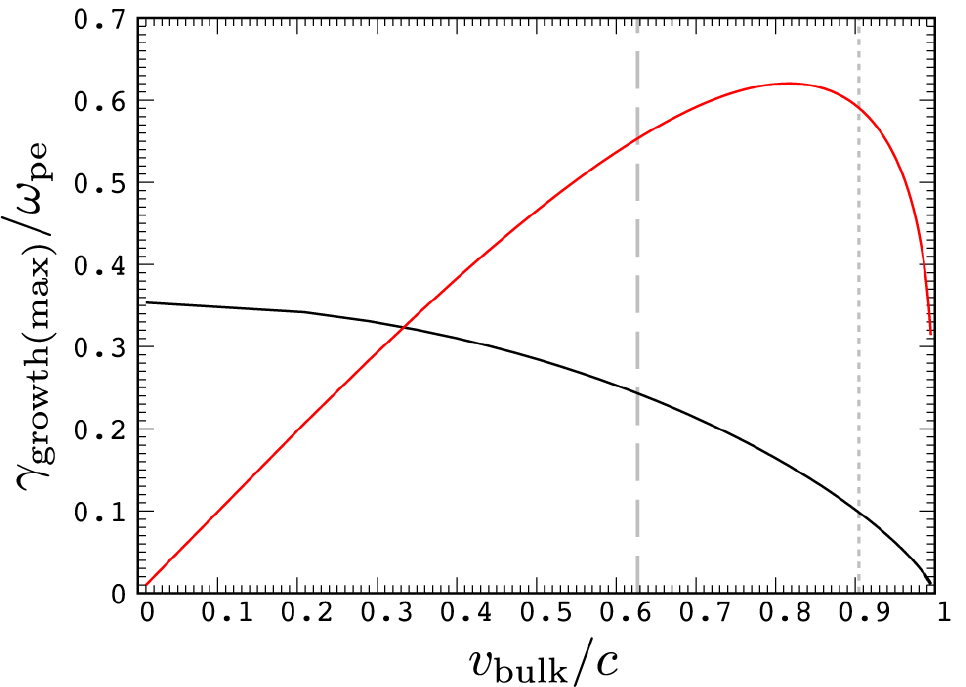} 
\caption{
Maximum linear growth rate of MI (red) and electron-scale Kelvin-Helmholtz instability (black) as a function of the bulk velocity of the counter plasma bulk flow.
The gray dashed and dotted lines show the bulk velocity of counter plasma bulk flow, which are obtained by the Lorentz transformation of our bulk velocity of jetted plasma $0.9 c$ (Lorentz factor $\simeq 2.3$) and $\simeq 0.99 c$ (Lorentz factor is 10), respectively.}
\label{fig:MI_dispersion}
\end{figure}

We demonstrate the maximum linear growth rate of the MI in the appendix.
Fist of all, the maximum growth rate of MI estimated for our simulations is shown in Figure \ref{fig:MI_dispersion} with including that of electron-scale Kelvin-Helmholtz instability for comparison, by following \cite{2015PhRvE..92b1101A}.
In both of our moderately and highly relativistic bulk velocity models, the bulk Lorentz factor is high enough that the MI dominates the electron-scale Kelvin-Helmholtz instability.
The linear analysis is carried out in the frame with zero-net-velocity in the whole simulation domain, i.e., the flow velocity has the same absolute value but opposite direction against the velocity-shear surface $+\bar{v}$ and $-\bar{v}$. 
The maximum growth rate of the MI is $\gamma_{\rm growth (max)}/\omega_{\rm pe} ~\sim~ (\bar{v}/c) \sqrt{\bar{\gamma}}$, where $\bar{\gamma}$ is the the Lorentz factor of $\bar{v}$. 
Our moderately relativistic bulk velocity model of $v_{\rm bulk} = 0.9 c$ corresponds to $\bar{v} ~\sim~0.6c$, where these are related by $v_{\rm bulk} = 2\bar{v}/(1+\bar{v}^2/c^2)$ and $\bar{\gamma} ~\simeq ~1.3$.
It should be noted that this dispersion relation is derived by assuming that the number density of the plasma is uniform in the zero-net-velocity frame, although in our simulations it is uniform in the rest frame of the plasma outside the velocity-shear surface and the plasma has moderately high thermal velocity ($0.1 c)$. For the moderately relativistic bulk velocity model, however, the relative velocity between these frame is ${\bar v} ~\sim ~0.6 c$ and the Lorentz factor is almost unity 
(${\bar \gamma} ~ \sim ~ 1.3$), so that the deviation is expected to be small enough in the context of the Lorentz contraction.\footnote{For the highly relativistic bulk velocity model, the discrepancy would be not negligible since the corresponding Lorentz factor is ${\bar \gamma} ~\sim~ 2.3$, i.e., the density inside and outside the velocity-shear surfaces are enhanced and reduced by the factor of $\sim ~ 2.3$.}
The plasma is also assumed to be cold in the linear analysis while our simulation assumed to be moderately warm (initial thermal velocity is set to be $0.1 c$). The finite thermal motion will slightly modify the maximum growth rate of our simulations but still agree with the linear analysis results as shown in Figure \ref{fig:time_energy}.

\end{document}